\documentstyle[art12]{article}
\def\t2{\tilde t}
\def\u2{\tilde u}
\def\s2{\tilde s}
\def\J{J/\psi}
\def\MJ{M_{\psi}}
\def\mc{m_c}
\def\cg{c(\gamma)g\to J/\psi c}
\def\gg{g(\gamma)g\to J/\psi g}

\def\sq{\sqrt{s_{\gamma p}}}
\def\pt{p_{\bot}}
\begin{document}
\begin{center}
 {\Large\bf Heavy Quarkonium Photoproduction at HERA\\
and Charm Content of the Photon\\}
\vspace{4mm}
V.A.Saleev\\
Samara State University,
Samara, 443011, Russia
\end{center}

\begin{abstract}
A study of the $\J$-meson photoproduction at large transverse momentum
at HERA energies via the charm quark excitation in
the photon is presented. Based on perturbative QCD,
nonrelativistic quark model and photon structure function formalism,
our calculation demonstrates that the
charm content of the photon may be very important for the $\J$
photoproduction at
the large transverse momentum. It is shown that at HERA energies
$\J$ production via the subprocess $\cg$ dominates over the resolved
photon contribution via the photon-gluon fusion at $\pt\ge 2$ GeV/c
and over the leading order direct photon-gluon fusion contribution  at
$\pt\geq 15$ GeV/c.
\end{abstract}

The  study  of        a  different  mechanisms of the
inclusive  $\J$  photoproduction  on  protons  is  of        a   particular
importance  because  the  process  of heavy quarkonium        production via
photon-gluon
fusion plays a crucial role in        the  measurement  of  gluon  structure
function in  a        proton        \cite{1,2,3,4}.  Previous calculations of
$\J$
photoproduction at   large   $\pt$   have  included  direct  charmonium
production via photon-gluon fusion  using  the        colour        singlet
model
in LO \cite{1,2} and NLO approximation \cite{5a},
diffractive  inelastic $\J$ production \cite{5},  resolved
photon contributions via subprocesses  $gg\to\J g$,
$gg\to\chi_c\to\J\gamma$ \cite{4} and $\J$ production from
$b$ quark  decays  \cite{6}.  Here  we        examine  the  diffractive-like
contribution of  the  charm quark excitation in a photon to
the $\J$ photoproduction at large $\pt$ via partonic subprocess $\cg$,
where charm  quarks  in  the  initial  state  are
generated by QCD evolution of  the  photon  structure
functions (PSF)  \cite{8}.  We        suppose  that  at  low $Q^2$ the charm
content of the photon is virtually nil,  but at $Q^2$ of order $m_c^2$
one has  sufficient  resolution to find charm quarks in photon.  Note,
that in the processes of the $\J$ photoproduction at large  $\pt$  the
relevant QCD scale $Q^2\sim \MJ^2+\pt^2>>m_c^2$.  Our calculation
is partly  the        same  as  the  approach  used  in   Ref.\cite{9}   for
description of        the  charm quark hadroproduction and based on the fact
that both the central and diffractive components of  charm  production
can be understood in the intrinsic charm structure function approach.
Recently we have
presented the results of the calculations  for        contributions  of  the
intrinsic charm  in  the  proton to the $\J$ photoproduction \cite{10}
and hadroproduction \cite{11}.

The study of a photon structure function \cite{8} in the
resolved-photon interaction at HERA energy and beyond is also very
interested \cite{12}. Usually it is supposed that at high energy in
the resolved-photon interaction the gluon content of a photon is
dominant. Howere, the heavy quark content of the photon is also an
increasingly important subject of study \cite{13}.
In this paper we would like to show that in the $\J$ photoproduction
at lager $\pt$ via
resolved-photon interaction the charm content of a photon is more
important and it is may be studied experimentally using $\J$ plus
open charm associated photoproduction.

The process of the $\J$ photoproduction on a proton in the resolved-photon
interaction is schematically presented in Fig. 1.

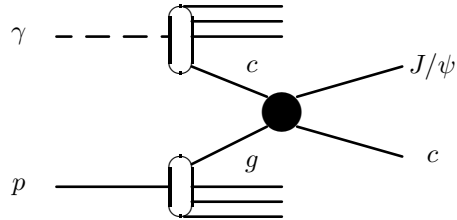
\def\emline#1#2#3#4#5#6{%
       \put(#1,#2){\special{em:moveto}}%
       \put(#4,#5){\special{em:lineto}}}
\unitlength=1mm
\special{em:linewidth 1pt}
\linethickness{1pt}
\begin{center}
\begin{picture}(60.00,34.00)
\emline{10.00}{10.00}{1}{25.00}{10.00}{2}
\put(26.50,10.00){\oval(3.00,8.00)[]}
\emline{10.00}{30.00}{3}{12.00}{30.00}{4}
\emline{14.00}{30.00}{5}{17.00}{30.00}{6}
\emline{19.00}{30.00}{7}{22.00}{30.00}{8}
\put(26.50,29.50){\oval(3.00,9.00)[]}
\emline{25.00}{30.00}{9}{24.00}{30.00}{10}
\emline{27.00}{34.00}{11}{40.00}{34.00}{12}
\emline{28.00}{32.00}{13}{40.00}{32.00}{14}
\emline{28.00}{30.00}{15}{40.00}{30.00}{16}
\emline{27.00}{6.00}{17}{40.00}{6.00}{18}
\emline{28.00}{8.00}{19}{40.00}{8.00}{20}
\emline{28.00}{10.00}{21}{40.00}{10.00}{22}
\put(40.00,20.00){\circle*{5.20}}
\emline{28.00}{26.00}{23}{38.00}{22.00}{24}
\emline{28.00}{13.00}{25}{38.00}{18.00}{26}
\emline{42.00}{22.00}{27}{56.00}{26.00}{28}
\emline{42.00}{18.00}{29}{56.00}{14.00}{30}
\put(5.00,30.00){\makebox(0,0)[cc]{$\gamma$}}
\put(5.00,10.00){\makebox(0,0)[cc]{$p$}}
\put(60.00,26.00){\makebox(0,0)[cc]{$\J$}}
\put(60.00,14.00){\makebox(0,0)[cc]{$c$}}
\put(36.00,26.00){\makebox(0,0)[cc]{$c$}}
\put(36.00,13.00){\makebox(0,0)[cc]{$g$}}
\end{picture}
\end{center}

\begin{figure}[h]
\vspace{2mm}
\caption{}
\end{figure}

The differential cross section for subprocess $\cg$
can be written as follows:
\begin{equation}
\frac{d\hat\sigma}{d\hat t}=\frac{\overline{|M|^2}}{16\pi(\hat s-\mc^2)^2}
\end{equation}
The explicit analytical formula for square of matrix element may be
found in Ref.\cite{11}.
In the framework of the conventional parton model
and photon structure function formalism the measurable cross-section
is obtained by folding the hard parton level cross-section  with
the respective parton densities:
\begin{eqnarray}
z\frac{d\sigma}{d^2\pt dz}&=&\frac{d\sigma}{d^2\pt dy^{\star}}=\int dx_1\int
dx_2
C_{\gamma}(x_1,Q^2)G_N(x_2,Q^2) \nonumber\\
&& \frac{d\hat\sigma}{d\hat t}
  (cg\to\J c)\frac{x_1x_2 s}{\pi}\delta(\hat s+\hat t+\hat u-\MJ^2
-2\mc^2).
\end{eqnarray}
Here: $\hat s=x_1x_2s+\mc^2,$ $\hat t=\MJ^2+\mc^2-x_1\sqrt{s}
M_{\bot}\exp(-y^{\star})$,
$\hat u=\MJ^2-x_2\sqrt{s} M_{\bot}\exp(y^{\star})$,
$M_{\bot}=\sqrt{\MJ^2+\pt^2},$
where $y^{\star}$ is the $\J$ rapidity, $\pt$ is the $\J$ transverse
momentum,
$z=(p_Np_{\psi})/(p_Np_{\gamma})=\frac{M_{\bot}}{\sqrt{s}}e^{y^{\star}}$,
$G_N(x_2,Q^2)$ is the gluon distribution function in a proton,
at the scale $Q^2=M_{\bot}^2$, $C_{\gamma}(x_1,Q^2)$ is the charm quark
distribution function in a photon, $s$ is the square of a total energy of
colliding particles in the $\gamma p$ center of mass reference frame.
We use in calculations LO GRV \cite{14} parameterization for
$G_N(x_2,Q^2)$ structure function.

The charm content of a photon can be presented as the composition of a
hadron--like PSF and a point--like (quark-gluon) PSF \cite{8}:
\begin{equation}
C_{\gamma}(x,Q^2)=C_{\gamma}^{had}(x,Q^2)+C_{\gamma}^{pl}(x,Q^2).
\end{equation}
In accordance with the vector-meson-dominance (VMD) model \cite{15},
the hadron--like $c$-quark PSF is presented via the charm quark
distribution function of $J/\psi$ meson:
\begin{equation}
C_{\gamma}^{had}(x,Q^2)= k\frac{4\pi\alpha}{f_{\psi}^2}C_{\psi}(x,Q^2)
\end{equation}
 with $1\leq k\leq 2$. The precise value of $ k$ clearly has
 to be extracted from experiment.  Similar to Ref.\cite{16}, where
  the photoproduction of charm hadrons
  in VMD model have been discussed,
 we used for $C_{\psi}(x,Q^2)$ the simple scaling parametrization,
which takes into account c-quark mass effects:
 \begin{equation}
  C_{\psi}(x)=49.5x^{2.2}(1-x)^{2.45}.
\end{equation}

The $c$-quark point-like part of the PSF can be calculated using
perturbative approach. As it was shown in Ref.\cite{15}, the distribution
of a heavy quark with mass $m_q$ in the photon near the threshold
($\sqrt{s_{\gamma g}}\geq 2m_q$) should be calculated from
the complete massive lowest order (Bethe - Heitler) cross section for
$\gamma^*\gamma\to q\bar q$ process:

\begin{equation}
C_{\gamma}^{pl}(x,Q^2)=\frac{3\alpha}{2\pi}e_c^2F(x,\frac{m_c^2}{Q^2}),
\end{equation}
where
\begin{equation}
F(x,r)=\beta[-1+8x(1-x)-4rx(1-x)]+
[x^2+(1-x)^2+4rx(1-3x)-8r^2x^2]\ln\biggl(\frac{1+\beta}{1-\beta}\biggr),
\end{equation}
$$\beta=\sqrt{1-4rx/(1-x)}.$$

Let us next consider the dynamical cutoff for the charm excitation
processes. For the typical charm excitation diagram, the charm
quark must receive sufficient transfer momentum, which is necessary
to excite $c\bar c$ pair. This implies a minimum dynamical resolution
$|\hat t|_{min}$ for the momentum transfer $\hat t$ of the $\cg$
subprocesses. We choose $|\hat t|_{min}=\mc^2$, although the
specification of the scale is uncertain by factors \cite{9}. The
dependence of the our results from the choice of $|\hat t|_{min}$ will
be discussed later.


Because of the relevant QCD scale $Q^2$ is order of $M_{\bot}$, we
consider at first $\pt$ distribution of the $\J$-mesons which are
generated in $\gamma p$ collisions via charm excitation in the photon.
Fig.~2 shows our predictions for $\J'$s $\pt$-spectra at the energy
$\sq=200$ GeV and $z<0.9$.
As the same Ref.\cite{9} we used in the calculation the next value of
the dynamical cutoff $|\hat t|_{min}=\mc^2$. Note, that $c-$quark PSF
contribution (curves 2 and 3 in Figs. 2-4) depends on $|\hat
t|_{min}$ only at low $\pt\leq 5$ Gev/c. So the charm excitation
mechanism (point-like PSF contribution) dominates over the LO direct
photon-gluon fusion at large transverse momentum $\pt\geq 15$ GeV/c
and more bigger the contribution of the gluon PSF at $\pt\geq 2$ GeV/c.

The values of the scaling
variable $z$ as well as transverse momentum can be used for separation
of the elastic ($z>0.9, \pt<1$ GeV/c) and inelastic $\J$
photoproduction events.
In the Fig.~3 the $z$-distributions for the $\J$ photoproduction at
$\sq=200$ GeV and $\pt>5$ GeV/c are shown.
In this region of $\pt$ our calculation are independent on the value of
$|\hat t|_{min}$. We see that at small $z\leq 0.4$ the contribution of
the $c-$quark PSF is equal to the direct photon-gluon fusion
contribution. The contribution of the hadron-like part of the
$c-$quark PSF, connected with the nonperturbative fluctuations
$\gamma\leftrightarrow J/\psi$,dominates over the point-like $c-$quark
PSF contribution at $z\geq 0.8$. The contribution of the gluon PSF is
very small at all $z\geq 0.1$.

In the Fig.~4 the results of calculation for the total cross section
of the $\J$ photoproduction at $z<0.9$ and $\pt>5$ GeV/c versus $\sq$
are presented.
The contribution of the subprocess $\cg$, which is independent of the
dynamical cutoff at large $\pt$, rapidly grows beginning with $\sq=40$
GeV and at $\sq=200$ GeV one has
$\sigma(\cg)/\sigma(\gamma g\to\J g)\approx 0.4$
for contribution of the point-like component and
$\approx 0.2$ for contribution of
the hadron-like component.
 At the energies $\sq>200$ GeV the behaviour of the total
cross sections both for the photon-gluon fusion as for the charm
excitation diagrams are the same and to be conditioned by the gluon
distribution function in the proton. The contribution of the resolved
photon interaction via subprocess $\gg$ in the total cross section at
large $\pt$ is two order of magnitude smaller the contributions of the
discussed above mechanisms.

Note, that we don't take into
consideration so-called $K$-factor which is needed as usual for
correct normalization of the leading order QCD predictions and experimental
data. In the present paper we accurately predict only the relative
contributions of the different $\J$ production mechanism at large $\pt$.
We have obtained that at high energies the total (perturbative and
nonperturbative) contribution of the $c-$quark PSF in the $J/\psi$
photoproduction at $\pt\geq 5$ GeV/c may be 50-60\% of the LO direct
photon-gluon fusion contribution.
We calculated here only direct $\J$ photoproduction cross section via
different mechanisms. It is well known that in the resolved photon
$\J$ photoproduction at large $\pt$  via gluon-gluon fusion \cite{4}
the two-step
mechanism dominates: $g(\gamma)g\to\chi_c g$ with decay
$\chi_c\to\J\gamma$ where $\chi_c$ are P-wave charmonium states.
That is why the calculation of the $\J$ photoproduction via subproceses
$c(\gamma)g\to\chi_c c$ with $\chi_c\to\J\gamma$ is also very
interesting. We are going to present the corresponding calculations in
the forthcoming papers. The another important source of the $\J$ with large
$\pt$ in $\gamma p$ interactions may be so-called c-quark
fragmentation into $\J$ or $\chi_c$ which was discussed recently
\cite{17}. However, the fragmentational contribution
connected with the c-quark, which was born in a hard partonic
interaction,  may be
separate experimentally from the contribution of the resolved photon
interactions \cite{12} via subprocesses $c(\gamma)g\to J/\psi c$ or
$c(\gamma)g\to \chi_c c$ which include as the fragmentational type $\J$
production as the recombinational one. Because
of the fact that in the resolved photon interactions it has
characteristic
hadron jet in the direction of the initial photon beam.

The discussed here $\J$ photoproduction mechanism via the
charm quark excitation can be used also for prediction of the open
charm production rates at high $\pt$ in $\gamma p$ collisions at
HERA energy and beyond \cite{18}.

{\it\bf Acknowledgements.}

  Author thank        A.~Likhoded, A.~Martynenko and N.~Zotov for useful
  discussions.
  This        research was supported by the Russian
  Foundation for Basic Research (Grant 93-02-3545).

\section*{Figure captions}
\begin{enumerate}
\item $J/\psi$ phtoproduction in the resolved photon interactions
\item
The $\pt$ distribution for $\J$ photoproduction at $\sq=200$ GeV
and all $z\leq 0.9$. The curve 1 is the direct photon-gluon fusion
contribution. The  curve 4 is the resolved photon
contribution via the $\gg$ subprocess. The curves 2 and 3 are the
contributions of the massive $c-$quark PSF, the
curve 2 corresponds to point-like part of the PSF, curve 3
corresponds to hadron-like part.
\item
The z-distribution for the $\J$ photoproduction at $\sq=200$ GeV
and $\pt>5$ GeV/c. Notation as in Fig.~2.
\item
The total cross section for the $\J$ photoproduction
 at $z<0.9 $ and $\pt>5$ GeV/c versus $\sq$. Notation as in Fig.~2.

\end{enumerate}


\begin{thebibliography}{99}
\bibitem{1} R.~Baier, R.~Ruckl, Nucl.Phys.B218~(1983)~289; B20~(1982)~1.
\vspace{-2.5mm}
\bibitem{2} E.L.~Berger, D.~Jones, Phys.Rev.D23~(1981)~1521;
\vspace{-2.5mm}
\bibitem{3} A.D.~Martin, C.-K.~Ng, W.J.~Stirling, Phys.Lett.B191~(1987)~200.
\vspace{-2.5mm}
\bibitem{4} H.~Jung, G.A.~Schuler, J.~Terron, Int.J.Mod.Phys.A32~(1992)~7955.
\vspace{-2.5mm}
\bibitem{5a} M.~Kraemer, Preprint DESY 95-155, 1995.
\vspace{-2.5mm}
\bibitem{5} G.A.~Schuler, J.~Terron, DESY preprint 92-017~(1992).
\vspace{-2.5mm}
\bibitem{6} Z.~Kunszt, Phys.Lett. B207 (1988) 103.
\vspace{-2.5mm}
\bibitem{8} E.~Witten, Nucl.Phys.B120~(1977)~189;\\
Ch.~Berger, W.~Wagner, Phys.Rep.146~(1987)~1.
\vspace{-2.5mm}
\bibitem{9} V.~Barger, F.~Halzen, W.Y.~Keung, Phys.Rev. D25 (1982) 112.
\vspace{-2.5mm}
\bibitem{10} V.A.~Saleev, Mod.Phys.Lett.A12~(1994)~1083.
\vspace{-2.5mm}
\bibitem{11} A.P.~Martynenko, V.A.~Saleev, Phys.Lett. B343 (1995) 381.
\vspace{-2.5mm}
\bibitem{12} W.~Buchmuller, G.~Ingelman, 1992 Physics at HERA,
Proc. Workshop 1991 (DESY).
\vspace{-2.5mm}
\bibitem{13} E.~Laenen et al., Phys.Rev. D49 (1994) 5753;\\
K.~Hagivara et al., Phys.Rev. D51 (1995) 3197;\\
T.~Sjostrand and G.~Schuler, Preprint CERN-TH/95-62.
\vspace{-2.5mm}
\bibitem{14} M.~Gluck, E.~Reya, A.~Vogt, Z.Phys. C53 (1992) 127.
\vspace{-2.5mm}
\bibitem{15} M.~Gluck, K.~Grassie, E.~Reya, Phys.Rev. D30 (1984) 1447.
\vspace{-2.5mm}
\bibitem{16} V.G.~Kartvelishvili, A.K.~Likhoded, V.A.~Petrov, Phys.Lett.B
(1978) 615;\\
A.K.~Likhoded, S.R.~Slabospitsky, A.N.~Tolstenkov, Yad.Fiz.38~(1982)~1240.
\vspace{-2.5mm}
\bibitem{17} E.~Braaten, M.A.~Doncheski, S.~Fleming, M.L.~Mangano,
Phys.Lett. B333 (1994) 548.
\vspace{-2.5mm}
\bibitem{18} M.~Cacciari, M.~Greco, Preprint DESY 95-103,1995;\\
M.Drees, R.~Godbole, Preprint MAD-PH-889, 1995;\\
B.A.Kniehl, M.~Kraemer, G.~Kramer, M.~Spira, Preprint DESY 95-098, 1995;\\
V.A.~Saleev, Preprint SSU-HEP 95-03, hep-ph 9505279.
\end{thebibliography}
\end{document}